\documentclass[11pt]{article}
\usepackage[letterpaper,margin=1in]{geometry}
\usepackage{amsmath,amssymb,booktabs,float,graphicx}
\usepackage{newtxtext,newtxmath}
\usepackage[numbers,sort&compress]{natbib}
\usepackage[hidelinks]{hyperref}

\graphicspath{{figures/}}
\newcommand{\E}{\boldsymbol E}
\newcommand{\Hh}{\boldsymbol H}
\newcommand{\neff}{n_{\mathrm{eff}}}
\newcommand{\ii}{\mathrm{i}}

\title{Does fluid pretraining help neural networks learn optical waveguide modes without solver labels?}
\hypersetup{pdftitle={Does fluid pretraining help neural networks learn optical waveguide modes without solver labels?},pdfauthor={Dorian Herle, Pedram Pad, Sathvik Bhagavan, Nadim Maamari}}
\author{Dorian Herle\textsuperscript{1,*}, Pedram Pad\textsuperscript{1},
Sathvik Bhagavan\textsuperscript{1,2} and Nadim Maamari\textsuperscript{1}}
\date{}
\begin{document}
\maketitle
\begin{center}
\small
\textsuperscript{1}CSEM SA, Rue Jaquet-Droz 1, 2002 Neuch\^atel, Switzerland\\
\textsuperscript{2}School of Computer and Communication Sciences, EPFL, 1015 Lausanne, Switzerland\\
\textsuperscript{*}\href{mailto:dorian.herle@csem.ch}{dorian.herle@csem.ch}
\end{center}

\begin{abstract}
Fluid pretraining improved optical waveguide mode prediction in a matched study. We trained Poseidon-L from fluid-pretrained or random weights under the same conditions. Maxwell constraints guided training and checkpoint selection without electromagnetic solver labels. On a test set of strip and ridge waveguides, mean normalized full-vector field errors were 14.3\% and 20.5\% with pretraining, compared with 32.4\% and 30.4\% for the matched random starts. Mean relative effective-index errors were 0.41--0.48\% and 1.25--1.34\%, respectively. After additional training, both pretrained models retained lower mean field and index errors across all three training settings on a second test set. With larger test regions in the Maxwell loss, the pretrained models reached mean field errors of 11.0\% and 12.8\%. These results show that fluid pretraining can help neural networks learn vector optical modes in the waveguide families tested here.
\end{abstract}

\section{Introduction}
A waveguide mode is defined by both its field shape and effective index. We ask whether starting from weights learned on fluid simulations improves these predictions when Poseidon-L is trained for guided modes under a fixed protocol. The inputs are cross-sectional permittivity images. Training uses Maxwell constraints without reference-field or effective-index targets. Pretrained and random initializations share the same architecture, training data, optimization settings and update budgets. We evaluate strip and ridge waveguide geometries excluded from training and checkpoint selection, at one wavelength, material pair and image scale.

Poseidon uses a multiscale transformer to learn fluid solution operators~\cite{herde2024}. Medvedev et al. demonstrated physics-informed adaptation to scalar Poisson and Helmholtz problems without downstream solution targets~\cite{medvedev2026}. Mi et al. used transfer learning for waveguide eigenmodes, adapting physics-informed models to similar structures~\cite{mi2026}. Normalization, orthogonality and eigenvalue selection are established tools in physics-informed eigenproblems~\cite{jin2022,hares2026}. Here, a geometry image produces quasi-TE and quasi-TM vector fields and indices in one forward pass. Our contribution is the comparison of pretrained and random starting weights under otherwise matched conditions, with electromagnetic training without solution labels. We then continue training each model with three loss settings: the original loss, larger local regions for checking the physics residual, and a weaker penalty encouraging larger effective indices. These experiments test whether the pretraining benefit persists and whether the loss changes improve prediction accuracy.

\section{Method}
\subsection{Image input and two vector outputs}
We consider lossless, nonmagnetic waveguides uniform along $z$, with fields proportional to $\exp(\ii\beta z-\ii\omega t)$. With $\beta=k_0\neff$ and $k_0=2\pi/\lambda_0$, the electric field obeys
\begin{equation}
\nabla\times(\nabla\times\E)=k_0^2\epsilon\E,
\qquad \nabla\cdot(\epsilon\E)=0,
\label{eq:maxwell}
\end{equation}
where $\epsilon$ is relative permittivity and $\partial_z=\ii\beta$. We fix $\lambda_0=1.55\,\mu\mathrm m$ and material indices 3.47 and 1.44. The input is a scalar $128\times128$ permittivity image over a $3\times3\,\mu\mathrm m^2$ domain. Geometry dimensions are used to draw the image, but are not separate neural inputs.

A small convolutional adapter maps the image to four channels for the original Poseidon-L ScOT backbone. The encoder, decoder, skip connections and native output layer are retained, with all weights trainable and the time input fixed to one. Two field heads return the real and imaginary parts of $E_x,E_y,E_z$ for quasi-TE and quasi-TM modes. These labels identify the dominant transverse electric component; both outputs retain all three components. Two index heads return effective indices bounded by the material indices. One forward pass gives both modes, without a solver or per-geometry optimization at inference. The full model has 628,578,902 parameters; the input adapter and output heads add 3378.

\begin{figure}[tbp]\centering\includegraphics[width=.86\linewidth]{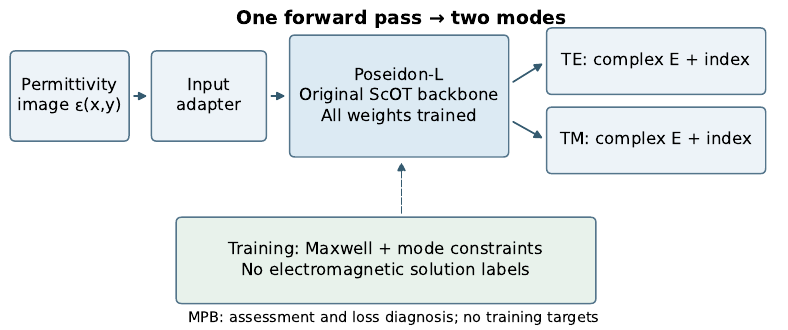}
\caption{Native Poseidon-L adaptation. The two modes share one backbone evaluation. Electromagnetic gradients and checkpoint selection use physics constraints; MPB supplies assessment and subsequent numerical diagnosis.}\label{fig:architecture}\end{figure}

\subsection{Label-free electromagnetic adaptation}
The network predicts a field and an effective index for each of the two modes. Training adjusts its weights to reduce violations of Maxwell's equations while encouraging the intended polarization, confinement and mode identity. No electromagnetic reference fields or indices are used as training targets. Before evaluating the loss, each predicted field is divided by its root-mean-square vector magnitude. This removes the arbitrary amplitude of an eigenmode; zero and nonfinite fields are rejected.

For one geometry, the loss averages the two individual-mode objectives and adds two penalties involving the mode pair:
\begin{equation}
\begin{aligned}
\mathcal L &= \frac12\sum_{p\in\{\mathrm{TE},\mathrm{TM}\}}\ell_p
             +L_{\mathrm{sep}}+L_{\mathrm{forward}},\\
\ell_p &= L_{M,p}+L_{G,p}+L_{\mathrm{pol},p}
          +L_{\mathrm{bnd},p}+L_{\mathrm{core},p}\\
       &\quad+L_{\mathrm{cons},p}+L_{\mathrm{valid},p}
          +L_{\mathrm{order},p}+L_{\mathrm{HF},p}.
\end{aligned}
\label{eq:trainingloss}
\end{equation}
Each term below includes its weight. We omit $p$ when a definition applies to both modes. The notation $[x]_+=\max(x,0)$ means that only positive violations contribute to a threshold penalty.

\noindent\begin{minipage}{\linewidth}
\paragraph{Maxwell residual.}
The residual $R_h$ measures how far the predicted field and index depart from the curl-curl equation in Eq.~\eqref{eq:maxwell}. For a local region centered at $q$ and a Cartesian component $j\in\{x,y,z\}$, it approximates
\begin{equation}
\begin{aligned}
(R_h)_{qj}={}&\int (\nabla_\beta\times\E)\cdot
 \overline{\nabla_\beta\times(\psi_q\boldsymbol e_j)}\,\mathrm dA\\
&-k_0^2\int\epsilon\E\cdot\overline{\psi_q\boldsymbol e_j}\,\mathrm dA.
\end{aligned}
\label{eq:weakresidual}
\end{equation}
Here $\psi_q$ is a smooth weighting function supported inside the local region, $\boldsymbol e_j$ is a unit vector, and $\nabla_\beta$ uses $\partial_z=\ii k_0\neff$. The subscript $h$ denotes evaluation as weighted sums on the image grid, with spacing $h=3/128\,\mu\mathrm m$. Each one-dimensional kernel factor samples the Wendland function~\cite{wendland1995} $(1-r)^4(4r+1)$ for $0\le r<1$ and zero elsewhere. The original loss uses $r=|j|/2$ at grid offsets $j=-2,\ldots,2$. The factors have unit discrete mass, and the derivative kernels reproduce affine derivatives. The two integrals should cancel for a Maxwell-consistent field: their difference is the local equation error. Variational physics-informed neural networks also use local weak residuals~\cite{kharazmi2021}. This weak form requires only first derivatives of the predicted field. The loss contribution is
\begin{equation}
L_M=r_M^2=
\frac{\langle|R_h(\E,\neff)|^2\rangle}
     {k_0^4\langle|\epsilon\E|^2\rangle}.
\label{eq:lossmaxwell}
\end{equation}
A larger residual means that the field and index are less consistent with the wave equation. The denominator makes the penalty dimensionless and independent of overall field amplitude. Numerators average over valid local test centers; denominators average over the full image.
\par
\end{minipage}\par\addvspace{0.8\baselineskip}

\noindent\begin{minipage}{\linewidth}
\paragraph{Gauss's law.}
In the source-free waveguide, the electric displacement $\epsilon\E$ must have zero divergence. A field can have a small curl-curl residual while still violating this additional condition. We therefore include
\begin{equation}
L_G=0.1r_G^2=
0.1\frac{\langle|G_h(\epsilon\E,\neff)|^2\rangle}
        {k_0^2\langle|\epsilon\E|^2\rangle},
\label{eq:lossgauss}
\end{equation}
where $G_h$ is the weak discrete divergence operator. The factor 0.1 is the fixed relative weight of this residual in the original objective.
\par
\end{minipage}\par\addvspace{0.8\baselineskip}

\noindent\begin{minipage}{\linewidth}
\paragraph{Polarization.}
For TE, define $\eta=\sum|E_x|^2/\sum(|E_x|^2+|E_y|^2)$; for TM, use the corresponding $E_y$ fraction. The term $L_{\mathrm{pol}}=10[0.55-\eta]_+^2$ penalizes a designated-component fraction below 55\% and is zero above it. The 55\% target provides a five-percentage-point margin above the 50\% boundary between the two polarization classes, while allowing mixed-component fields. It is a fixed training heuristic, not a physical constant; sensitivity to this threshold was not assessed. All three complex field components are retained.
\par
\end{minipage}\par\addvspace{0.8\baselineskip}

\noindent\begin{minipage}{\linewidth}
\paragraph{Boundary confinement.}
Let $f_b$ be the fraction of summed $|\E|^2$ in the outer eight pixels of the image. The term $L_{\mathrm{bnd}}=[f_b/0.005-1]_+^2$ is zero up to a boundary fraction of 0.5\% and penalizes larger values. It favors fields contained within the finite image. This is a soft confinement prior, not an exact open-boundary condition.
\par
\end{minipage}\par\addvspace{0.8\baselineskip}

\noindent\begin{minipage}{\linewidth}
\paragraph{Core participation.}
Define $f_c=\sum f|\E|^2/\sum|\E|^2$, where $f=(\epsilon-1.44^2)/(3.47^2-1.44^2)$ is the core-fill image. The term $L_{\mathrm{core}}=[(0.1-f_c)/0.1]_+^2$ penalizes participation below 10\%. It discourages solutions concentrated outside the guiding region. Both $f_b$ and $f_c$ use unweighted squared electric-field magnitude, rather than optical power or material-weighted energy.
\par
\end{minipage}\par\addvspace{0.8\baselineskip}

\noindent\begin{minipage}{\linewidth}
\paragraph{Index consistency.}
The model predicts the field and index separately. A second index estimate $n_*$ comes from the predicted field: projecting $R_h$ onto that field gives a quadratic in index, whose larger, numerically guarded root is used. With $\Delta n=3.47-1.44$, $L_{\mathrm{cons}}=[(\neff-n_*)/\Delta n]^2$ penalizes disagreement between the two estimates relative to the material-index range. Neither estimate uses a reference solution.
\par
\end{minipage}\par\addvspace{0.8\baselineskip}

\noindent\begin{minipage}{\linewidth}
\paragraph{Index validity.}
For the quadratic projection, let $\E_c$ be the field on the valid test centers and define
\begin{equation*}
P(n)=\frac{\Re\langle\E_c,R_h(\E,n)\rangle}{k_0^2\langle|\E_c|^2\rangle}=an^2+bn+c .
\end{equation*}
Evaluations at $n=0,\pm1$ give $c=P(0)$, $b=[P(1)-P(-1)]/2$ and $a=[P(1)+P(-1)]/2-c$.
We use $\widetilde a=\max(a,10^{-4})$, $d=b^2-4\widetilde a c$ and $n_*=[-b+\sqrt{\max(d,10^{-6})}]/(2\widetilde a)$.
The validity term is
\begin{equation}
\begin{aligned}
L_{\mathrm{valid}}={}&[0.05-a]_+^2+[10^{-6}-d]_+^2\\
&+\left(\frac{[1.44-n_*]_+}{\Delta n}\right)^2
 +\left(\frac{[n_*-3.47]_+}{\Delta n}\right)^2.
\end{aligned}
\label{eq:lossvalidity}
\end{equation}
The first two terms penalize coefficients or discriminants below the numerical acceptance levels; the last two penalize an index outside the cladding--core interval. A root is accepted only if $a>0.05$, $d>10^{-6}$ and $1.44<n_*<3.47$. Arithmetic clamping keeps calculations finite but does not make an invalid root acceptable.
\par
\end{minipage}\par\addvspace{0.8\baselineskip}

\noindent\begin{minipage}{\linewidth}
\paragraph{Index ordering.}
Several guided modes may satisfy Maxwell's equations. To favor the larger-index mode within each designated polarization class, we use
\begin{equation}
L_{\mathrm{order}}=w_{\mathrm{ord}}\begin{cases}
(3.47-n_*)/\Delta n,&\text{accepted root},\\
1,&\text{otherwise},
\end{cases}
\label{eq:lossindex}
\end{equation}
with $w_{\mathrm{ord}}=0.2$ in the original loss. For an accepted root, increasing $n_*$ reduces this contribution. An invalid root receives a constant cost rather than a reward. This is a mode-selection bias; it does not prove that the fundamental mode has been found.
\par
\end{minipage}\par\addvspace{0.8\baselineskip}

\noindent\begin{minipage}{\linewidth}
\paragraph{High-frequency regularization.}
Set $L_{\mathrm{HF}}=f_{\mathrm{HF}}$, where $f_{\mathrm{HF}}$ is the fraction of electric Fourier intensity above 0.25 cycles per pixel in either transverse direction. This term has weight one. It discourages rapid pixel-to-pixel oscillations that the discrete residual can miss. It is a numerical regularizer, rather than another Maxwell equation.
\par
\end{minipage}\par\addvspace{0.8\baselineskip}

\noindent\begin{minipage}{\linewidth}
\paragraph{Mode separation.}
Polarization penalties alone do not establish modal separation. We also use the symmetrized flux product of the two predicted modes~\cite{meepmodes},
\begin{equation}
\begin{aligned}
S_{ij}&=\left\langle[\overline{\E_i}\times\Hh_j
             +\E_j\times\overline{\Hh_i}]_z\right\rangle,\\
L_{\mathrm{sep}}&=\frac{|S_{01}|^2}{\max(S_{00}S_{11},10^{-16})}.
\end{aligned}
\label{eq:losspair}
\end{equation}
Here $i,j\in\{0,1\}$ identify the two modes, and $\Hh=(\nabla\times\E)/(\ii k_0)$ is derived by Faraday's law in the adopted units. This product involves both electric and magnetic fields. The penalty discourages a large cross-mode flux overlap; its denominator is guarded for finite arithmetic. On arbitrary neural fields the ratio need not lie between zero and one and is not clipped to that interval.
\par
\end{minipage}\par\addvspace{0.8\baselineskip}

\noindent\begin{minipage}{\linewidth}
\paragraph{Forward propagation.}
The separation term does not by itself require either mode to carry forward power. We therefore add
\begin{equation}
L_{\mathrm{forward}}=\frac{10}{2}\sum_{p\in\{\mathrm{TE},\mathrm{TM}\}}
\left[0.05-\frac{S_{pp}}{2\cdot3.47}\right]_+^2.
\label{eq:lossforward}
\end{equation}
Each mode is penalized when its normalized self-flux falls below 0.05. The factor $1/2$ averages the two modes and the factor 10 sets this penalty's weight. This complements the cross-mode separation constraint by discouraging backward or insufficient forward self-flux.
\par
\end{minipage}\par\addvspace{0.8\baselineskip}

The complete loss guides the network toward plausible guided modes, but a small discrete loss does not certify field accuracy or fundamental-mode identity. We assess the predictions against independently computed MPB modes.

For each of two seeds, pretrained and random Poseidon models have identical adapter initialization and receive the same sequence of images. Each trains for 5000 updates on 384 images, using one image per update and no augmentation. Backbone and adapter learning rates are $3\times10^{-5}$ and $3\times10^{-4}$. Twelve separate development images determine checkpoint selection every 250 updates: fixed physics guards establish eligibility, then development loss ranks eligible checkpoints. No electromagnetic labels, teacher outputs or MPB accuracy scores enter gradients or checkpoint selection. ``Label-free'' refers to electromagnetic adaptation; the public backbone was pretrained with fluid solution data.

\subsection{Independent assessment}
The initial assessment contains 32 strips and 32 ridges excluded from training and development. Strip widths and heights span $0.45$--$0.95$ and $0.32$--$0.55\,\mu\mathrm m$. Ridge widths, heights above the slab and slab thicknesses span $0.45$--$1.00$, $0.20$--$0.38$ and $0.12$--$0.20\,\mu\mathrm m$. Duplicate images and near-identical earlier geometries are excluded. This tests interpolation within two families at a fixed wavelength, material pair and image scale.

MPB~\cite{johnson2001} computes reference modes in a periodic $3\,\mu\mathrm m$ transverse supercell at 48 pixels/$\mu\mathrm m$. Fixed-frequency roots are checked by fresh deterministic eight-band solves, with relative frequency error at most $10^{-5}$. Among these bands, the largest guided index in each polarization class defines the target. The full-domain transverse intensity ratio $\sum|E_x|^2/(\sum|E_x|^2+\sum|E_y|^2)$ separates TE ($\ge0.5$) from TM. References are sampled at neural image centers. Resolution checks increase the grid from 48 to 96 pixels/$\mu\mathrm m$ on eight predetermined geometries per test set. Additional domain checks diagnose ridge 025 after assessment. All primary scores use the original references.

We use two accuracy measures: normalized full-vector field error $e_E$ and relative effective-index error $e_n$. With $\widehat{\E}=\E/\|\E\|_2$,
\begin{equation}
e_E=\min_{\phi\in\mathbb R}\|e^{\ii\phi}\widehat{\E}_{\rm NN}-\widehat{\E}_{\rm ref}\|_2,
\qquad e_n=\frac{|n_{\rm NN}-n_{\rm ref}|}{n_{\rm ref}}.
\end{equation}
Each complete three-component field first receives unit norm, then one common phase is aligned. The score measures the remaining vector-amplitude difference: 10\% means a difference norm of 0.10, not 10\% power loss. It does not divide by an individual component or pixel value. Near-zero values therefore cannot create a singular relative error; regions where both fields are small contribute little. A spurious predicted component still counts when the reference component is small. Magnitude images combine the components and discard phase, so they can conceal vector differences. At the fixed wavelength, $e_n$ also measures relative propagation-constant error; device phase accuracy depends on propagation length.

Tables average per-mode errors over all 64 geometries and both polarizations, keeping seeds separate. Fig.~\ref{fig:errorcomparisonmain} shows mean field errors and every individual geometry separately for TE and TM. Overlap is equivalent to $e_E$ for each mode and adds no independent accuracy information. Paired case-bootstrap intervals use 10,000 resamples stratified by geometry family, retaining TE/TM and seed pairs. They quantify geometry sampling conditional on the fitted models, not variation across training runs.

\section{Results}
\subsection{Matched initialization comparison}
Both pretrained models reduce field and index errors relative to their matched random initializations after 5000 updates (Table~\ref{tab:focusedinitial}). The two-seed mean reduction in field error is 13.99 percentage points, with a paired case-bootstrap 95\% interval of [13.71, 14.23]. The corresponding reduction in index error is 0.851 percentage points [0.606, 1.149]. These narrow intervals describe the sampled geometries for the two fitted seed pairs. All four selected checkpoints are at update 5000; no seed is discarded.
\begin{table}[H]\centering\small
\caption{Initial test set after 5000 updates. Each mean includes all 64 geometries and both polarizations (128 predictions per seed). Lower errors indicate better agreement; Fig.~\ref{fig:errorcomparisonmain} separates TE and TM.}\label{tab:focusedinitial}
\begin{tabular}{lrr}\toprule
Initialization & Mean $e_E$ (\%) & Mean $e_n$ (\%)\\\midrule
Pretrained 1 & 14.31 & 0.480 \\
Pretrained 2 & 20.50 & 0.409 \\
Random 1 & 32.35 & 1.343 \\
Random 2 & 30.43 & 1.247 \\
\bottomrule\end{tabular}\end{table}

\begin{figure}[H]\centering\includegraphics[width=.88\linewidth]{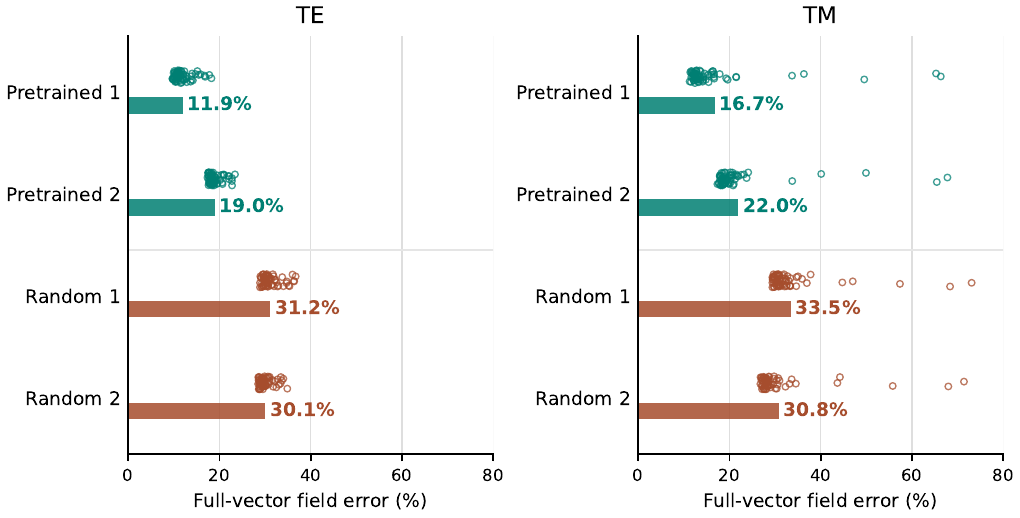}
\caption{Field errors on the initial test set after 5000 matched updates. Bars and numerical labels give the mean field error; each dot represents one of the 64 geometries. Errors use all three complex electric-field components. TE and TM use the same horizontal scale, and lower errors indicate better agreement. Small vertical offsets separate the dots visually and have no numerical meaning. Seeds 1 and 2 identify paired pretrained/random training runs. Every case is shown.}\label{fig:errorcomparisonmain}\end{figure}

Fig.~\ref{fig:errorcomparisonmain} compares field errors separately for TE and TM. Each bar gives the mean over 64 geometries, and each dot gives the error for one geometry. Shorter bars and dots farther left indicate better field agreement. The pretrained models have lower mean errors in both polarizations and both seeds, while the dots at large TM errors show the cases that remain difficult.

\subsection{Additional training evaluated on a second test set}
Each selected 5000-update model is copied into three runs, each receiving up to 2500 more updates. The settings are ``original loss'' (Eq.~\eqref{eq:trainingloss}); ``larger test regions'' (radius two to four pixels for the Maxwell and Gauss residuals and the field-derived index); and ``weaker index preference'' ($w_{\mathrm{ord}}$ reduced from 0.2 to 0.02). The first assessment motivated these changes.

All runs reset Adam's accumulated gradient statistics and reduce both learning rates tenfold. Changes from the starting model therefore combine more updates, the reset and lower learning rates. Comparisons among the three settings share this schedule. Common physics guards determine checkpoint eligibility; each setting's development loss ranks eligible checkpoints.

A separate 64-geometry test set is reserved before checkpoint selection, and weights are fixed before its MPB evaluation. Table~\ref{tab:focusedcontinuation} compares starting and continued models on this same set. Reference fields and indices never enter training or checkpoint selection; comparisons between the two test sets do not measure training improvement.
\begin{table}[H]\centering\small
\caption{Second test set: models before and after additional training. Each cell lists seed 1, seed 2; each mean includes all 64 geometries and both polarizations (128 predictions). Starting models have 5000 updates. Each additional-training setting allows 2500 more updates, with physics-only checkpoint selection.}\label{tab:focusedcontinuation}
\begin{tabular}{@{}llrr@{}}\toprule
Training setting & Initialization & Mean $e_E$ (\%) & Mean $e_n$ (\%)\\\midrule
Before additional training & Pretrained & 13.68, 19.83 & 0.449, 0.404 \\
Before additional training & Random & 31.85, 29.90 & 1.396, 1.292 \\
Original loss & Pretrained & 12.50, 15.65 & 0.348, 0.327 \\
Original loss & Random & 29.66, 27.18 & 1.130, 1.051 \\
Larger test regions & Pretrained & 11.02, 12.82 & 0.367, 0.317 \\
Larger test regions & Random & 28.49, 25.64 & 1.116, 1.039 \\
Weaker index preference & Pretrained & 12.15, 14.97 & 0.342, 0.326 \\
Weaker index preference & Random & 29.30, 26.93 & 1.108, 1.043 \\
\bottomrule\end{tabular}\end{table}

With the original loss, pretrained mean field errors fall from 13.68\% to 12.50\% and from 19.83\% to 15.65\% for seeds 1 and 2. Larger test regions lower them further to 11.02\% and 12.82\%, with a small index-error increase for seed 1 and decrease for seed 2. Pretrained models outperform their matched random starts in both mean errors under all three settings.

\subsection{Limits of field accuracy}
The field images show the total magnitude $|\E|=\sqrt{|E_x|^2+|E_y|^2+|E_z|^2}$. Combining the components into one magnitude hides how the field is divided among directions and discards their phases. Two images can therefore look similar even when the underlying vector fields differ substantially. For ridge 025 TM in the second test set, the two models trained with larger test regions have field errors of 39.49\% and 36.94\%, despite similar total-magnitude images and index errors below 1\%.

\begin{figure}[!t]\centering\includegraphics[width=.88\linewidth]{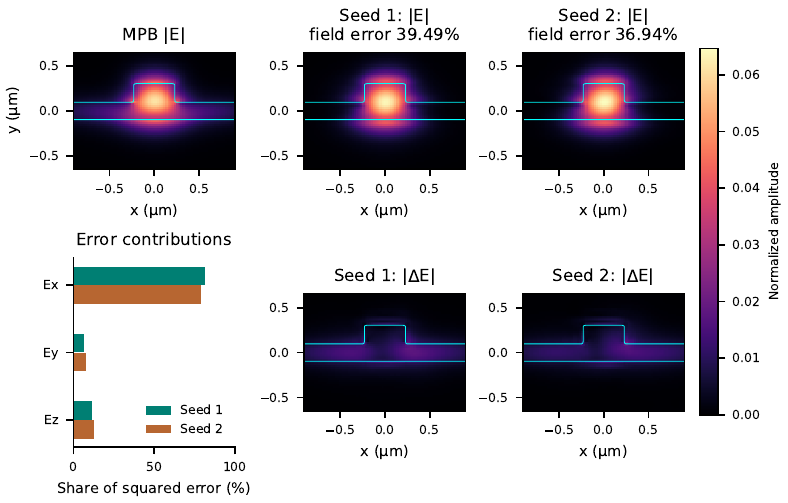}
\caption{Ridge 025 TM from the second test set, for models trained with larger test regions. Top: similar total field magnitudes can conceal a substantial component mismatch. Bottom: about 80\% of the squared prediction error comes from $E_x$, alongside maps of $|\Delta\E|$ after full-vector normalization and one common phase alignment. All maps share an amplitude scale; dim differences extending over many pixels can accumulate. Error integrates the full domain; the central window is displayed. This failure was selected after assessment.}\label{fig:failuremain}\end{figure}

The main discrepancy is an underestimated $E_x$ component. In the MPB reference, $E_x$ contributes 17.03\% of the summed squared electric-field magnitude over the full image. Its contribution is only 1.63\% for seed 1 and 2.84\% for seed 2. These percentages describe the balance between field components. By contrast, 81.67\% and 79.13\% describe the share of the total squared prediction error caused by the $E_x$ mismatch (Fig.~\ref{fig:failuremain}). A component that carries about 17\% of the reference squared magnitude can thus account for about 80\% of the error when the model substantially underestimates it. Neither percentage is an optical-power fraction; both diagnose the same full-vector error rather than adding new accuracy measures.

The component fractions concern squared magnitudes, so their square roots give the component norms after the complete vector is normalized to one. The reference $E_x$ norm is $\sqrt{0.1703}\simeq0.413$, compared with $\sqrt{0.0163}\simeq0.128$ and $\sqrt{0.0284}\simeq0.169$ in the predictions. Even if the $E_x$ shapes and phases matched perfectly, these norm differences alone would require full-vector field errors of at least 28.5\% and 24.4\%. Similar total-magnitude images cannot remove this missing component amplitude.

The periodic reference domain also matters: neighboring copies of the geometry can affect field tails. For ridge 025 TM, the phase-aligned difference between normalized MPB fields is 8.81\% when the supercell grows from 3 to $4\,\mu\mathrm m$ and 0.0081\% when only the eigensolver tolerance tightens. Extended checks reach 192 pixels/$\mu\mathrm m$ and a 10-$\mu\mathrm m$ supercell, meeting the recorded adjacent-domain and resolution sensitivity targets for both polarizations. The component discrepancy persists, with neural field errors 51.30\%/49.09\%. Separately, all matched comparisons retain the mean pretraining advantage against finer-grid references on eight predetermined geometries per cohort. Primary scores retain the original finite reference problem; these checks do not certify continuum accuracy or a noise floor for small components.

\section{Discussion and conclusion}
Fluid pretraining lowers mean field and index errors in both seeds and both test sets under the matched protocol. This result applies to the tested strip and ridge families, wavelength, material pair, image scale and training budget. Two seeds give limited evidence about variation between training runs. The study does not explain the transfer mechanism or establish optimization convergence or broader generalization. Different training settings could change the comparison.

The achieved accuracy remains limited. Larger local test regions improve pretrained field agreement, with a seed-dependent index trade-off and a persistent TM component error. Domain-sensitive MPB checks limit absolute accuracy: the reported errors measure agreement with specified finite references. Low discrete losses, small index errors and similar magnitude images do not certify vector-mode accuracy. Interface discretization, boundary assumptions and the Fourier regularizer remain unresolved.

We also trained compact U-Net~\cite{ronneberger2015} controls under the same update budget. One U-Net seed has lower mean field error with far fewer parameters, but higher mean index error. Model and computational superiority remain unestablished. More training seeds and domain refinement across predetermined cohort subsets are needed for stronger claims.

\section*{Reproducibility}
Reference fields and indices were kept separate from training and checkpoint selection for both test sets. We reloaded and verified the selected inference checkpoints.

\bibliographystyle{unsrtnat}
\bibliography{references}
\end{document}